\documentclass[conference]{IEEEtran}
\usepackage[T1]{fontenc}
\usepackage[utf8]{inputenc}
\usepackage{amsmath,amsfonts,amsthm,bm}
\usepackage{cuted}
\usepackage{color}
\makeatletter
\newcommand{\mathleft}{\@fleqntrue\@mathmargin0pt}
\newcommand{\mathcenter}{\@fleqnfalse}
\makeatother
\IEEEoverridecommandlockouts
\usepackage{cite}
\usepackage{amsmath,amssymb,amsfonts}
\usepackage{algorithmic}
\usepackage{graphicx}
\usepackage{textcomp}
\def\BibTeX{{\rm B\kern-.05em{\sc i\kern-.025em b}\kern-.08em
\kern-.1667em\lower.7ex\hbox{E}\kern-.125emX}}

\begin{document}

\title{Active fault tolerant control for twin wind turbine subject to asymmetric fault}

\author{\IEEEauthorblockN{Mariem Makni}
\IEEEauthorblockA{\textit{LA2MP, ENIS, Sfax-Tunisie} \\
\textit{QUARTZ, ENSEA, Cergy-Pontoise}\\
France\\
mariem.makni@ensea.fr}
\and
\IEEEauthorblockN{Ihab Haidar}
\IEEEauthorblockA{\textit{ENSEA} \\
\textit{QUARTZ}\\
Cergy-Pontoise, France \\
ihab.haidar@ensea.fr}
\and
\IEEEauthorblockN{Jean-Pierre Barbot\hspace{1cm}}
\IEEEauthorblockA{\textit{LS2N, UMR 6004, CNRS, Nantes\hspace{1cm}} \\
\textit{QUARTZ, ENSEA, Cergy-Pontoise \hspace{1cm}}\\
France\hspace{1cm} \\
barbot@ensea.fr\hspace{1cm}}
\and
\IEEEauthorblockN{\hspace{2cm}Franck Plestan}
\IEEEauthorblockA{\textit{\hspace{2cm}Ecole Centrale de Nantes} \\
\textit{\hspace{2cm}LS2N, UMR 6004, CNRS}\\
\hspace{2cm}Nantes, France\\
\hspace{2cm}franck.plestan@ec-nantes.fr}
\and
\IEEEauthorblockN{Nabih Feki}
\IEEEauthorblockA{\textit{University of Sousse} \\
\textit{HIAST}\\
Sousse, Tunisie \\
fekinabih@gmail.com}
\and
\IEEEauthorblockN{Mohamed Slim Abbes}
\IEEEauthorblockA{\textit{ENIS} \\
\textit{LA2MP}\\
Sfax, Tunisie \\
Slim.Abbes@enis.tn}
}

\maketitle

\begin{abstract}
                                           
This paper addresses the problem of control of a twin wind turbine which is subject to an electrical fault affecting only one stator phase of one turbine. An active fault tolerant control is proposed. The performance and robustness of the proposed control, comparing to a passive fault tolerant one developed in the literature, are shown through numerical simulations.

\end{abstract}

\section{Introduction}
The concept of Twin Wind Turbine (TWT), patented in \cite{cc1}, which is composed of two
identical turbines is subject of this paper. The main advantage of this structure is the fact that rotation is ensured without the need of a dedicated actuator but by creating difference between drag force of each turbine.
The objective of control strategies applicated to the twin wind turbine is to control the whole structure position by keeping it in front of the wind direction at all times, in order to optimize energy production whereas the second objective is to control the electrical generator. Knowing that, electrical machines can be inevitably be damaged or afflicted by numerous failures as well as eccentricities, short winding fault, sensor faults and permanent magnet demagnetization fault \cite{cc2}, \cite{cc3}, an active fault tolerant control should be considered. Indeed, as shown in \cite{cc4}, the fault tolerance of the healthy control strategy developed in \cite{cc4} is very questionable when considering some electrical faults. Some works have been performed on turbine control.\\
In \cite{cc4}, the healthy control used in \cite{cc5} has been tested with taking into consideration an insulation fault in the
Permanent Magnet Synchronous Machine (PMSM) of only one turbine while keeping the second turbine without any defect. Nonetheless, the used control after its robustification (see~\cite{cc4} for more details) was not sufficient and its fault tolerant capabilities was limited.
In \cite{cc6}, observer based on Fault Detection and Isolation FDI for wind turbines is proposed and extended in \cite{c2} in which accommodation schemes and fault detection of benchmark model were be presented. Focus on pitch system fault, a Kalman-Bucy filter based diagnosis was developed in \cite{c3} to detect faults relating to blade sensors. Both active and passive fault tolerant controllers indeed a robust controller are designed in \cite{c4} in order to accommodate the pitch fault and to compensate aerodynamic model uncertainties respectively. Vidal et al. \cite{c5} develops a fault tolerant control of pitch actuators in wind turbine to handle parameter variations and to robust the pitch system under faults.\\
In this work, an active fault tolerant control of the twin wind turbine is investigated in presence of an asymmetric fault which affect one stator phase of one turbine. As the inter-turns short circuit is the most common fault, this case will be considered here. Due to this fault, the symmetry of the three stator phases of the PMSM is not yet been adequately respected. A classical control based on the symmetries is totally ineffective for asymmetric faults greater than $7\%$. Consequently, the control model is done in $abc$-frame instead of $dq$-frame in order to stabilize the imbalance caused by this fault and allow the structure to continue operating until the presence of the defect.\\
The paper is organized as follows. In section II, after description of the twin wind turbine dynamics, specific attentions of electrical part are done for both healthy and faulty dynamics in $abc$-frame. In the next section, an active fault tolerant control is presented. This control is based on $abc$-frame in order to take into account the asymmetry effect. In section IV, the proposed active fault tolerant control is compared with a passive fault tolerant control. The simulation results highlight the interest of the proposed method. Conclusion and  open problems for future research are drawn in the section V.

\section{Problem statement}

\subsection{Healthy twin wind turbine description}
\begin{figure}[thpb]
      \centering
      \includegraphics[scale=0.75]{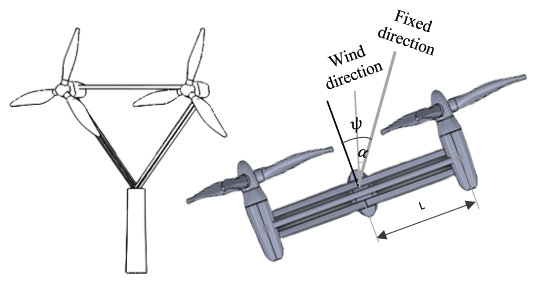}
      \caption{SEREO concept - face (left) and top (right) view \cite{cc4}}
      \label{psialpha}
   \end{figure}

In this paper, a specific concept of wind turbine is considered \cite{cc1}. As shown in Fig. \ref{psialpha}., the twin wind turbine is composed of two turbines mounted on the same tower. As mentioned above, the orientation in front of wind is ensured without need an actuator. Furthermore, to keep this orientation while maintain optimal power production requires a robust control. More details on twin wind turbine description can be found in \cite{c7}.\\
Mechanical powers from kinetic energy and aerodynamic torques applied to the rotor by the wind are respectively given by:
\begin{align}
&{P_{i}} = \frac{\pi \rho}{2} {C_{pi}}\left( {{\lambda _i},{\beta _i}} \right){R_p^2}{\left( {{{V_v}}\cos (\psi  - \alpha )} \right)^3}
\label{p}
\\
&{\Gamma _{a_{i}}} =\frac{{{P_i}}}{{{\Omega _i}}}= \frac{\pi \rho}{{2{\lambda _i}}} {C_{pi}}\left( {{\lambda _i},{\beta _i}} \right){R_p^3}{\left( {{{V_v}}\cos (\psi  - \alpha )} \right)^2}
\end{align}
where $i=\left\{ {1,2} \right\}$ denotes the first and the second turbines, ${V_v}$ is the wind speed and $R_p$ is the radius of blade, $\rho$ is the air density. The angle $\alpha$ represents the angle between the wind direction and the fixed direction and the angle $\psi$ is defined as the angle between horizontal axis of structure and the fixed direction.  The power coefficient ${C_{pi}}$  \cite{c12}, characterized by nonlinear functions depend on ${\lambda_i}$, the tip speed ratio and ${\beta_i}$, the pitch angle of blade. The tip speed ratio is proportional to the rotor angular speed, and reads as:
\begin{align}
{\lambda _i} = \frac{{R_p{\Omega _i}}}{{{V_v}\cos \left( {\psi  - \alpha } \right)}}\label{lamda}
\end{align}
To extract the maximum of wind energy, the twin wind turbine must be oriented face to wind and this is realized when the angle of orientation $\psi$ attain its reference $\alpha$. It can be argued that this orientation is the common input of the two turbines.\\
This orientation is made by the torque extracted from the difference between forces generated by the two turbines which enables system rotation around its vertical axis. Thus, dynamics of system rotation is:
\begin{align}
{d_r}\ddot \psi  =  - {f_r}\dot \psi  + \left( {{F_1} - {F_2}} \right)l \label{psieq}
\end{align}
where $d_r$ is the inertia moment, $f_r$ is the friction coefficient associated to yaw motion, $l$ is the distance between the horizontal and vertical axis of structure and ${F_1} - {F_2}$ is the difference between the two drag forces. The drag force is defined by:
\begin{align}
{F_{i}} = \frac{\pi \rho}{2} {C_{di}}\left( {{\lambda _i},{\beta _i}} \right){R_p^2}{\left( {{{V_v}}\cos (\psi  - \alpha )} \right)^2}
\label{ff}
\end{align}
Among objectives, optimal power production has to be achieved. For that, power coefficients 
$C_{pi}^{opt}$ which correspond to $\beta _i^{opt}$ and $\lambda _i^{opt}$ \cite{c12}, have to reach their optimal values.
Otherwise, in this concept, adjusting the pitch angles is addressing not only to maximize power but also to enable the rotation of the system (drag force of each turbine depend on its pitch blade (\ref{ff})). \\
The twin wind turbine is equipped with two salient permanent magnet synchronous machines. The models of both safe and faulty machines are elaborated in detail.
For each turbine, the three-phase permanent magnet synchronous machine model in coordinates $abc$ is represented by the following equations:
\begin{align}
V_i = \left( {{r_s} + \frac{{d {{L^s}({\theta _{ei}})}}}{{dt}}} \right)I_i + {{L^s}({\theta _{ei}})} \frac{d}{{dt}}I_i + \frac{d}{{dt}}{E_{mi}}
\label{PMSMh}
\end{align}
where $V_i = {\left[ {\begin{array}{*{20}{l}}
{{v_{a{n_i}}}}&{{v_{b{n_i}}}}&{{v_{c{n_i}}}}
\end{array}} \right]^T}$ and $I_i = {\left[ {\begin{array}{*{20}{c}} {{i_{a_i}}}&{{i_{b_i}}}&{{i_{c_i}}}
\end{array}} \right]^T}$ are respectively the voltage and current vectors of the three phases. The stator resistance and the inductance matrix containing self inductance of each winding, and the mutual inductance are respectively $r_s$ and $L^s(\theta_{ei})$. Due to saliency, this matrix depends on the rotor electrical angular position $\theta_{ei}$. The electromotive force vector
$E_{mi}={\left[ {\begin{array}{*{20}{l}} {{e_{m{a_i}}}}&{{e_{m{b_i}}}}&{{e_{m{c_i}}}} \end{array}} \right]^T}$ depends on the flux linkage $\phi_f$ produced by the permanent magnet. Its elements can be done by the following functions:
\begin{flalign*}
&e_{m{a_i}} = \phi_f \cos(\theta _{ei})\\\nonumber
&e_{m{b_i}} = \phi_f \cos(\theta _{ei}-\frac{2\pi}{3})\\\nonumber
&e_{m{c_i}} = \phi_f \cos(\theta _{ei}+\frac{2\pi}{3}).
\end{flalign*}

\subsection{Faulty twin wind turbine description}
The robustness of control based on $dq$-frame with a fault which affects only one turbine, on performances of the TWT has been investigated in \cite{cc4}. The validity domain of the proposed control was limited. In this paper, an active fault-tolerant control in $abc$-frame is developed for largest validity domain because the $dq$-transformation can no longer be used due to the strong electrical asymmetry. So, for faulty case, $abc$-model is considered.\\For sake of simple presentation, inter-turns short circuit failure in the $b$-phase is considered. Remark that the model can be easily adjusted on $a$ and $c$ phases by index permutation. Recognizing that the fault persists in only one turbine, the index $i$ here will denote the faulty turbine 1 or 2.
\begin{align}
\nonumber&{V_i} = diag\left( {\begin{array}{*{20}{l}}
{{r_s}}~{\left( {1 - \bar \mu } \right){r_s}}~{{r_s}}
\end{array}} \right){I_i} + \frac{{d{{L^f}({\theta _{ei}},\bar \mu)}}}{{dt}}{I_i} \\
& + {{L^f}({\theta _{ei}},\bar \mu)} \frac{d}{{dt}}{I_i} + \frac{d}{{dt}}\left[ {\begin{array}{*{20}{l}}
{{e_{m{a_i}}}}~{\left( {1 - \bar \mu } \right){e_{m{b_i}}}}~{{e_{m{c_i}}}}
\end{array}} \right]^T
\end{align}
Notice that the severity $\bar \mu$ can be defined as the factor of the short-circuited turns $N_{bd}$ devising by the total number of $b$-phase winding $N_{bt}$. ${{L^f}({\theta _{ei}},\bar \mu)}$ is the fault inductance matrix expressed by (\ref{Lff}). For a healthy machine, the inductance has the same expression of the fault inductance with respect to $\bar \mu=0$.
\begin{align}
&{{{L^{f}}\left( {{\theta _{ei}}},\bar \mu \right)}} = \label{Lff}
\\\nonumber
&\left[ {\begin{array}{*{20}{l}}
{{L_a}\left( {{\theta _{ei}}} \right)}&{\left( {1 - \bar \mu } \right){M_{ab}}\left( {{\theta _{ei}}} \right)}&{{M_{ac}}\left( {{\theta _{ei}}} \right)}\\
{\left( {1 - \bar \mu } \right){M_{ba}}\left( {{\theta _{ei}}} \right)}&{\left( {1 - \bar \mu } \right){L_b}\left( {{\theta _{ei}}} \right)}&{\left( {1 - \bar \mu } \right){M_{bc}}\left( {{\theta _{ei}}} \right)}\\
{{M_{ca}}\left( {{\theta _{ei}}} \right)}&{\left( {1 - \bar \mu } \right){M_{cb}}\left( {{\theta _{ei}}} \right)}&{{L_c}\left( {{\theta _{ei}}} \right)}
\end{array}} \right]
\end{align}
Self inductance of each winding are the diagonal elements $(L_a, L_b, L_c)$, and the rest of elements design the mutual inductance between different phase winding $(M_{ab}, M_{ac}, M_{ba}, M_{bc}, M_{ca}, M_{cb})$. Elements of the above inductance matrix are highly nonlinear and depend on ${\theta _{ei}}$, which may the applicability of $abc$-phase model comes at the expense of model complexity.\\
With respect to this model, in the next section, an active fault-tolerant control will be proposed in order to broaden the validity domain of fault severity. Remark that, the passive fault-tolerant control used in \cite{cc4} is only efficient on $\bar \mu  \in \left[ {0..7\% } \right]$.

\section{Active fault-tolerant control}

\begin{figure}[thpb]
      \centering
      \includegraphics[scale=0.6]{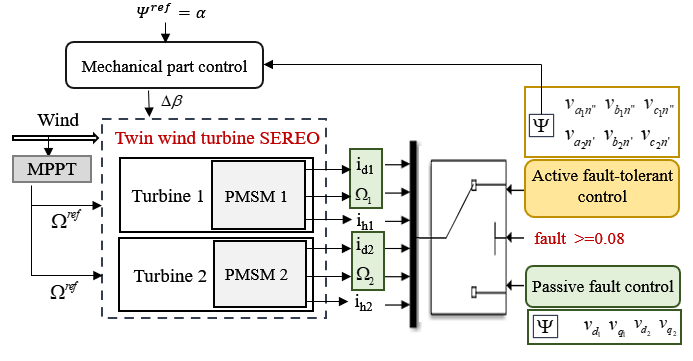}
      \caption{Control scheme of the twin wind turbine}
      \label{control}
   \end{figure}

This section focus on the active fault tolerant control. Comparing to a passive fault tolerant control, in this paper the assumption of symmetric phases is not supposed due to the presence of the fault. Consequently, due to this asymmetry, it is not relevant to design a control in the $dq$-frame. This approach leads to two additional states in the state vector ($i_{a_1}, i_{b_1}, i_{c_1}, i_{a_2}, i_{b_2}, i_{c_2}$ three-phase currents are considered instead of two-phase currents $i_{d_1}, i_{q_1}, i_{d_2}, i_{q_2}$) and in the same way, two extra inputs are introduced ($v_{an_1}, v_{bn_1}, v_{cn_1}, v_{an_2}, v_{bn_2}, v_{cn_2}$ instead of $v_{d_1}, v_{q_1}, v_{d_2}, v_{q_2}$). Therefore, in this work, an extended nonlinear system based on $abc$-frame is given by:
\begin{align}
\dot x = {f_{\bar \mu }}(x,t) + {g_{\bar \mu }}(x,t)u
\end{align}
with ${f_{\bar \mu }}(x,t)$ and ${g_{\bar \mu }}(x,t)$ are respectively the vector and the matrix given in the appendix, the state and the input vectors can be written as:
\begin{align}
\nonumber &x = {\left[ {\begin{array}{*{20}{l}}
{{\beta _1}}~{{\beta _2}}~\psi~{\dot \psi}~{{i_{a1}}}~{{i_{b1}}}~{{i_{c1}}}~{{\Omega _1}}~{{i_{a2}}}~{{i_{b2}}}~{{i_{c2}}}~{{\Omega _2}}
\end{array}} \right]^T}\\\nonumber
&u = {\left[ {\begin{array}{*{20}{l}}
{\Delta \beta }~{{v_{a{n_1}}}}~{{v_{b{n_1}}}}~{{v_{c{n_1}}}}~{{v_{a{n_2}}}}~{{v_{b{n_2}}}}~{{v_{c{n_2}}}}
\end{array}} \right]^T}
\end{align}
Figure \ref{control}. shows two possibilities of control. These strategies take into account the phase in which the defect occurs, so they can be adapted for the healthy case. Nevertheless, the passive fault control after robustification is limited to fault smaller to $8\%$ whereas the active fault tolerant control is efficient for severity largely higher than $8\%$.
The objectives of the control is to compensate the effect of the fault on electrical part with respect to the mechanical part. Consequently, from the mechanical point view, the faulty machine must have the same behavior of the healthy machine. For that, a well-driven electrical torque is assigned only with $i_{q_i}$ where the direct current $i_{d_i}$ and the homopolar component $i_{hi}= \frac{1}{3} ({i_{a_i}+i_{b_i}+i_{c_i}})$ are assigned to zero in order to avoid ripple effects on electromagnetic torque. Regarding that the angular velocity depends on electromagnetic torque (\ref{omega}) which is proportional to the quadratic current $i_{q_i}$, it might not be very useful to control directly $i_{q_i}$. The electromagnetic torque is given by:
\begin{align}
&{\Gamma _{em_{i}}} = p\left( {{L_d} - {L_q}} \right){i_{di}}{i_{qi}} + p{\phi _f}{i_{qi}}
\label{omega}
\end{align}
where $L_d$, $L_q$ are $dq$-axis inductance and $p$ is the pole-pair number. As it is mentioned above, for the non-symmetric phases case, the direct current indeed the homopolar component $i_{hi}$ must be controlled to zero. This latter doesn't be considered on the control based on $dq$-rotating frame. So, the difference between the two control strategies is that to consider $7$ outputs instead $5$ in \cite{cc4}. Outputs of control are then choosing as $y$:
\begin{align}
y = h(x) = \left[ {\begin{array}{*{20}{l}}
{\psi  - \alpha }\\
{{i_{d1}}}\\
{{\Omega _1} - \Omega _1^{ref}}\\
{{i_{h1}}}\\
{{i_{d2}}}\\
{{\Omega _2} - \Omega _2^{ref}}\\
{{i_{h2}}}
\end{array}} \right]
\end{align}

The successive derivatives of the outputs lead to the following equation:

\begin{align}
{{{ {{y^{({\varepsilon})}}} }} = {{{\Lambda (x,t)} }} + {{{\Theta(x,t)}}}{{ {u} }}}
\end{align}

where $y^{({\varepsilon})}=(y_1^{(3)}, y_2^{(1)}, y_3^{(2)}, y_4^{(1)}, y_5^{(1)},y_6^{(2)},y_7^{(1)})^{T}$, with ${y^{(i)}=\frac{{d^i}{y}}{dt^i}}$, for $i\geq 1$.\\
Note that ${\varepsilon}=(\varepsilon_1, \varepsilon_2, \varepsilon_3, \varepsilon_4, \varepsilon_5,\varepsilon_6, \varepsilon_7)=(3,1,2,1,1,2,1)$ are the relative degrees of $(y_1, y_2, y_3, y_4, y_5, y_6, y_7)$ respectively. The vector field $\Lambda$ and the decoupling matrix $\Theta$,  identifying in the appendix, correspond to the successive derivatives of $h_i$.\\
Since $\Theta$ is a regular matrix, the following
decoupling control based on a new control $\bar z$ while respecting homogeneity arguments could be obtained as:

\begin{equation}
{{{{u} }} = {\Theta(x,t)} }^{ - 1}\left( {{{ {{\bar z}} }} - {{ {\Lambda(x,t)} }}} \right)
\label{y}
\end{equation}

\begin{equation}
\bar z  = \left[ {\begin{array}{*{20}{l}}
{ - {K_\psi }{{\left| {{\sigma _{\varepsilon \psi }}} \right|}^{{\mu _{1,3}}}}sign\left( {{\sigma _{\varepsilon \psi }}} \right)}\\
{ - {K_{{\Omega _1}}}{{\left| {{\sigma _{\varepsilon {\Omega _1}}}} \right|}^{{\mu _{4,5}}}}sign\left( {{\sigma _{\varepsilon {\Omega _1}}}} \right)}\\
{ - {K_{{i_d}_1}}{{\left| {{\sigma _{\varepsilon {i_d}_1}}} \right|}^{{\mu _{6,6}}}}sign\left( {{\sigma _{\varepsilon {i_d}_1}}} \right)}\\
{ - {K_{{i_h}_1}}{{\left| {{\sigma _{\varepsilon {i_h}_1}}} \right|}^{{\mu _{7,7}}}}sign\left( {{\sigma _{\varepsilon {i_h}_1}}} \right)}\\
{ - {K_{{\Omega _2}}}{{\left| {{\sigma _{\varepsilon {\Omega _2}}}} \right|}^{{\mu _{8,9}}}}sign\left( {{\sigma _{\varepsilon {\Omega _2}}}} \right)}\\
{ - {K_{{i_d}_2}}{{\left| {{\sigma _{\varepsilon {i_d}_2}}} \right|}^{{\mu _{10,10}}}}sign\left( {{\sigma _{\varepsilon {i_d}_2}}} \right)}\\
{ - {K_{{i_{h\partial }}_2}}{{\left| {{\sigma _{\varepsilon {i_h}_2}}} \right|}^{{\mu _{11,11}}}}sign\left( {{\sigma _{\varepsilon {i_h}_2}}} \right)}
\end{array}} \right]
\label{vartheta}
\end{equation}

The expressions of $\sigma _{\varepsilon \psi }, \sigma _{\varepsilon {\Omega _1}}, \sigma _{\varepsilon i{d_1}}, \sigma _{\varepsilon i{h_1}} \sigma _{\varepsilon {\Omega _2}}$, $\sigma _{\varepsilon i{d_2}}$, $\sigma _{\varepsilon i{h_2}}$ can be calculated based on \cite{cc5} and the different homogeneity degrees $\mu _{i,j}$ is given by:
\mathleft
\begin{align}
\nonumber
&\mu_{i,j}=\max\{1-\delta\sum_{l=i}^{j}\frac{|z_l|}{|z_l|+\varepsilon_i},0\}, \qquad 
\\&\text{for}\quad 1\leq i,j\leq 11,\quad \delta  > 1, \quad {\varepsilon _i} > 0
\end{align}

Knowing that $\sum\limits_{k = 1}^7 {{\varepsilon_k}}  = 11<12$, so there is a zero dynamics of dimension 1.
\mathcenter
\begin{equation}
{\dot z_{12}} = \frac {{2}{\beta^{ref}}}{t_{\beta}}-{\frac {1}{t_{\beta}}{z_{12}}}.
\label{z10}
\end{equation}
Comparing with \cite{cc4}, the dynamics of zero $z_{12}$ is even input-to-state stable (input: $\beta_{ref}$) \cite{c13}. From equation (\ref{z10}), the dynamic of zero depends on the external input which is not directly influenced by the defect.\\
The advantage of the active fault-tolerant control, that it use the estimation of the fault. This estimation may be done by many diagnosis methods for example \cite{c14}, \cite{c15}. Nevertheless, estimation of fault is not the topic of this paper.

\section{Simulation results and discussion}

\begin{figure}[thpb]
      \centering
      \includegraphics[scale=0.65]{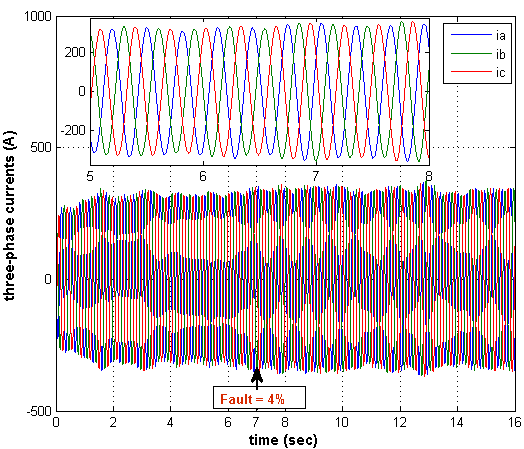}
      \caption{Effects of the fault on the three-phase currents}
      \label{iabcrc}
   \end{figure}

\begin{figure}[thpb]
      \centering
      \includegraphics[scale=0.7]{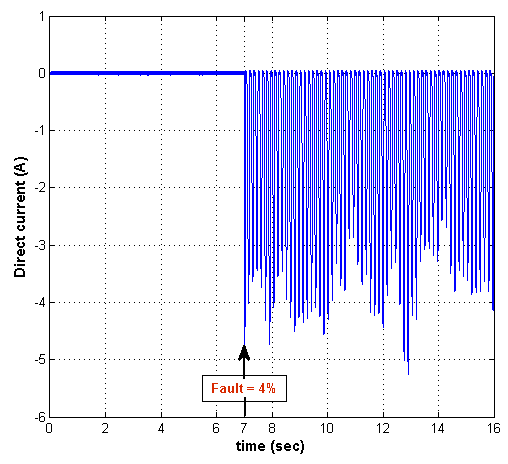}
      \caption{Direct current waveform related to the asymmetric phases}
      \label{idrc}
   \end{figure}

\begin{figure}[thpb]
      \centering
      \includegraphics[scale=0.7]{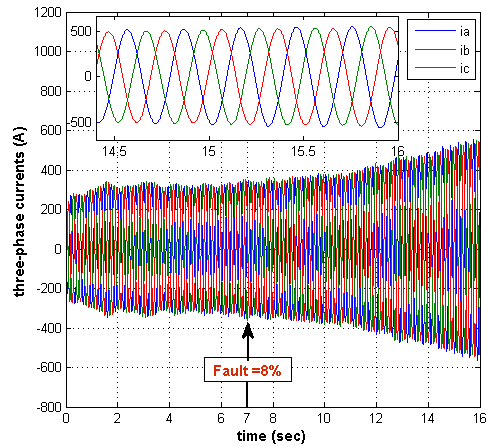}
      \caption{Effects of the fault on the three-phase currents}
      \label{iabcf}
   \end{figure}

In the case of a classical control, the voltage references are generated in terms of the $dq$-rotating frame \cite{c16}, \cite{c17} and then transformed back into $abc$-frame. However, ones the fault occurs only on one phase, the three-phases currents increase and generate an uncontrolled direct current $i_{d_i}$.\\
In order to stabilize these currents,  the gains related to the direct currents $K_{i_{d1}}$ and $K_{i_{d2}}$ in equation \eqref{vartheta} are increased (''robustified''), as it was proposed in \cite{cc4}.\\ As it can be seen in Fig. \ref{iabcrc}. and Fig. \ref{idrc}., the fault occurs at the time $t = 7s$ with the severity of $4 \%$. This recent ''robustification", gives a satisfactory behavior with respect to minor faults but it is not efficient if the defect exceeds $8 \%$ (Fig. \ref{iabcf}.).\\

As it was described in section III, more tougher faults can not be consider by classical control. In the next, simulation results of the proposed active fault tolerant control are presented. Based on the fact that an inter-turn short circuit  on the stator $b$-phase persists with an important severity (equal to $20 \%$), it seems necessary to use the $abc$-model for the control design.\\
   
   \begin{figure}[thpb]
      \centering
      \includegraphics[scale=0.5]{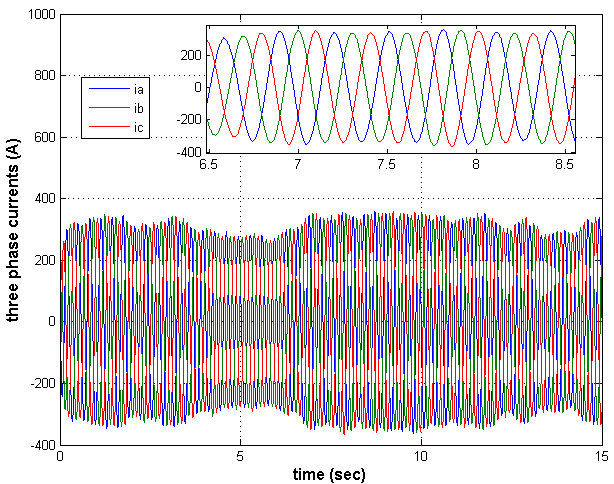}
      \caption{The three-phase currents after active control}
      \label{iabcac}
   \end{figure}
   
   \begin{figure}[thpb]
      \centering
      \includegraphics[scale=0.5]{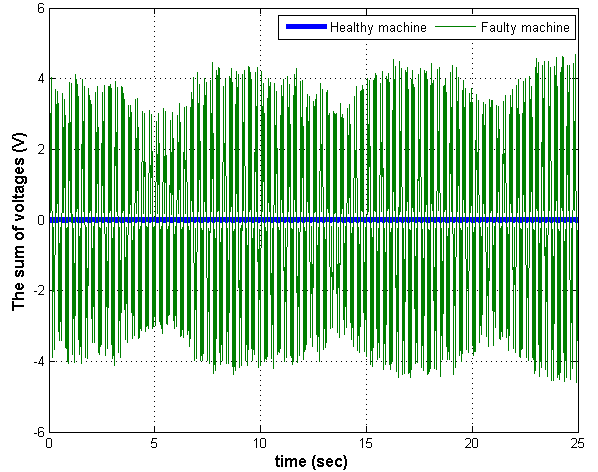}
      \caption{Comparison between the sums of voltages for healthy and faulty machines}
      \label{sumV}
   \end{figure}
   
   \begin{figure}[thpb]
      \centering
      \includegraphics[scale=0.5]{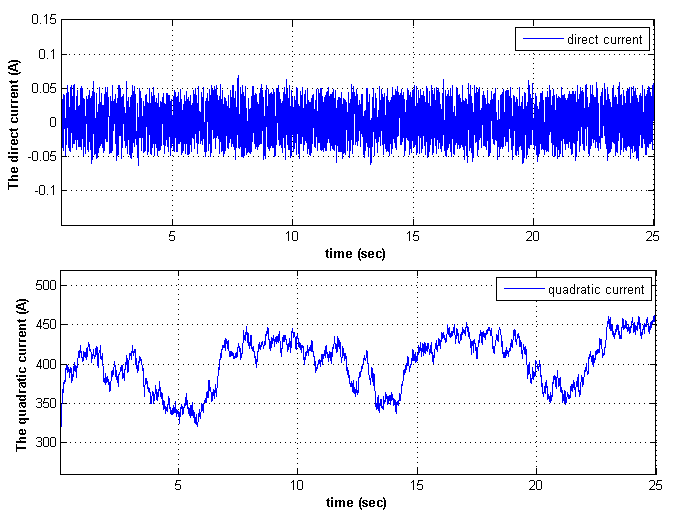}
      \caption{The direct and the quadratic currents after control}
      \label{idq}
   \end{figure}
   
   \begin{figure}[thpb]
      \centering
      \includegraphics[scale=0.5]{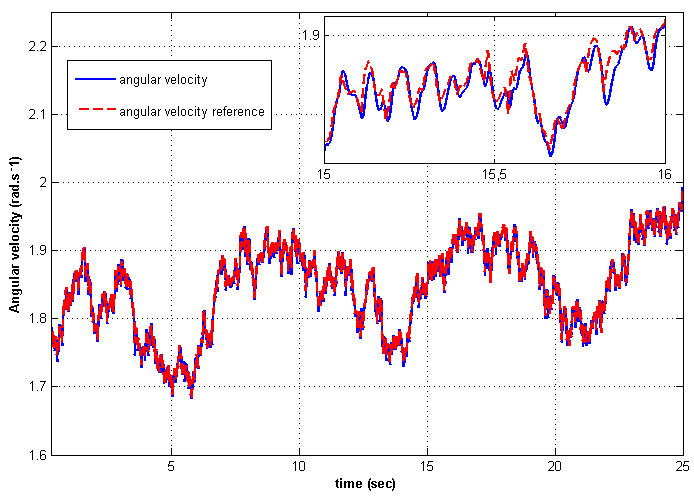}
      \caption{The controlled angular velocity and its reference waveform}
      \label{angularspeed}
   \end{figure}
   
   \begin{figure}[thpb]
      \centering
      \includegraphics[scale=0.5]{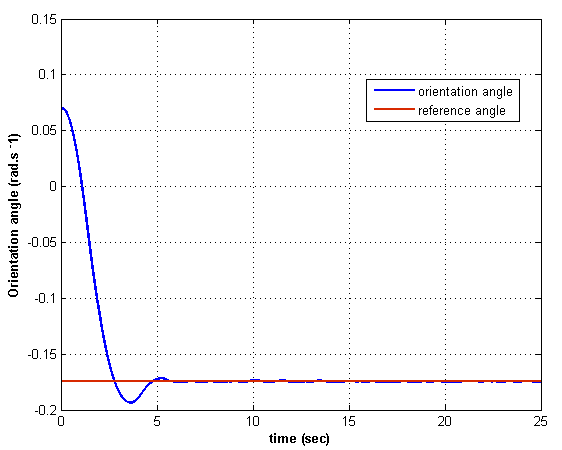}
      \caption{The Angle of orientation after control}
      \label{angularorientation}
   \end{figure}
   
Figure Fig. \ref{iabcac} shows the three-phase currents of the faulty machine after active fault tolerant control. Despite of asymmetric phases in the PMSM under faulty conditions, the sum of the three-phase currents are equal to zero. This reflects the fact that the direct current and the homo-polar component are controlled to be equal to zero (Fig. \ref{idq}.).\\Thus, from Figures Fig. \ref{angularspeed}. and Fig. \ref{angularorientation}, it appears that the angular speed as well as the angle orientation follow their references. These highlight that all control objectives are achieved even if the fault is sever.

\section{Conclusion}

In this paper, it was shown that with respect to asymmetric electrical faults, it is necessary to use the $abc$-frame. Consequently, an active fault-tolerant control strategy for a twin wind turbine was proposed in this frame. This control is tested for both healthy and faulty cases, when the presence or not of fault and its severity are well estimated. Simulation results highlight the well founded of the proposed approach.\\
In future research, an estimation method of the fault based on sparsity assumptions \cite{c19} coupled with the proposed active fault-tolerant control will be investigated.
The main problem of active fault-tolerant control based on faults estimation  process, is its global stability. Because, even if, the faults estimation process is stable and the active fault-tolerant control is also stable, these do not guarantee that the coupling of both is stable.

\setcounter{equation}{0}
\numberwithin{equation}{section}
\begin{strip}
\appendix
Considering that the fault occurs in the electrical machine of the first turbine, the model can be given by the equation \eqref{fannex}, where $L^f$ and $L^s$ present the inductance matrix on the faulty and the healthy cases respectively. These matrix depend on the electrical angular position $\theta_{ei}$.
\mathleft
\begin{equation}
{f_{\bar \mu }}(x,t) = \left[ {\begin{array}{*{20}{l}}
{ - \frac{{{\beta _1}}}{{{T_\beta }}}}\\
{ - \frac{{{\beta _2}}}{{{T_\beta }}}}\\
{\dot \psi }\\
{ - \frac{{{f_r}}}{{{d_r}}}\dot \psi  + \left( {{F_1} - {F_2}} \right)L}\\
{ - \left[ {\begin{array}{*{20}{l}}
{L^f_{11}}^{ - 1}&{L^f_{12}}^{ - 1}&{L^f_{13}}^{ - 1}
\end{array}} \right]\left( {{\left[ {\begin{array}{*{20}{l}}
{{r_s}}&{(1 - \bar \mu ){r_s}}&{{r_s}}
\end{array}} \right]}{I_1} + \frac{d}{{dt}} {{L^f}} {I_1} + \frac{{d{E_{m1}}}}{{dt}}} \right)}\\
{ - \left[ {\begin{array}{*{20}{l}}
{L^f_{21}}^{ - 1}&{L^f_{22}}^{ - 1}&{L^f_{23}}^{ - 1}
\end{array}} \right]\left( {{\left[ {\begin{array}{*{20}{l}}
{{r_s}}&{(1 - \bar \mu ){r_s}}&{{r_s}}
\end{array}} \right]}{I_1} + \frac{d}{{dt}} {{L^f}} {I_1} + \frac{{d{E_{m1}}}}{{dt}}} \right)}\\
{ - \left[ {\begin{array}{*{20}{l}}
{L^f_{31}}^{ - 1}&{L^f_{32}}^{ - 1}&{L^f_{33}}^{ - 1}
\end{array}} \right]\left( {{\left[ {\begin{array}{*{20}{l}}
{{r_s}}&{(1 - \bar \mu ){r_s}}&{{r_s}}
\end{array}} \right]}{I_1} + \frac{d}{{dt}} {{L^f}} {I_1} + \frac{{d{E_{m1}}}}{{dt}}} \right)}\\
{\frac{1}{J}\left( {{\Gamma _{{a_1}}} - {\Gamma _{e{m_1}}} - {f_v}{\Omega _1}} \right)}\\
{ - \left[ {\begin{array}{*{20}{l}}
{L^s_{11}}^{ - 1}&{L^s_{12}}^{ - 1}&{L^s_{13}}^{ - 1}
\end{array}} \right]\left( {{r_s}{I_2} + \frac{d}{{dt}} {{L^s}}{I_2} + \frac{{d{E_{m2}}}}{{dt}}} \right)}\\
{ - \left[ {\begin{array}{*{20}{l}}
{L^s_{21}}^{ - 1}&{L^s_{22}}^{ - 1}&{L^s_{23}}^{ - 1}
\end{array}} \right]\left( {{r_s}{I_2} + \frac{d}{{dt}} {{L^s}} {I_2} + \frac{{d{E_{m2}}}}{{dt}}} \right)}\\
{ - \left[ {\begin{array}{*{20}{l}}
{L^s_{31}}^{ - 1}&{L^s_{32}}^{ - 1}&{L^s_{33}}^{ - 1}
\end{array}} \right]\left( {{r_s}{I_2} + \frac{d}{{dt}} {{L^s}} {I_2} + \frac{{d{E_{m2}}}}{{dt}}} \right)}\\
{\frac{1}{J}\left( {{\Gamma _{{a_2}}} - {\Gamma _{e{m_2}}} - {f_v}{\Omega _2}} \right)}
\end{array}} \right]
\label{fannex}
\end{equation}
\mathleft
\begin{equation}
{g_{\bar \mu }}(x,t) = \left[ {\begin{array}{*{20}{c}}
{\frac{1}{{{T_\beta }}}}&0&0&0&0&0&0\\
{ - \frac{1}{{{T_\beta }}}}&0&0&0&0&0&0\\
0&0&0&0&0&0&0\\
0&0&0&0&0&0&0\\
0&{L^f_{11}}^{ - 1}&{L^f_{12}}^{ - 1}&{L^f_{13}}^{ - 1}&0&0&0\\
0&{L^f_{21}}^{ - 1}&{L^f_{22}}^{ - 1}&{L^f_{23}}^{ - 1}&0&0&0\\
0&{L^f_{31}}^{ - 1}&{L^f_{32}}^{ - 1}&{L^f_{33}}^{ - 1}&0&0&0\\
0&0&0&0&0&0&0\\
0&0&0&0&{L^s_{11}}^{ - 1}&{L^s_{12}}^{ - 1}&{L^s_{13}}^{ - 1}\\
0&0&0&0&{L^s_{21}}^{ - 1}&{L^s_{22}}^{ - 1}&{L^s_{23}}^{ - 1}\\
0&0&0&0&{L^s_{31}}^{ - 1}&{L^s_{32}}^{ - 1}&{L^s_{33}}^{ - 1}\\
0&0&0&0&0&0&0
\end{array}} \right]
\end{equation}
For a simplified representation, considering that:\\
\mathleft
\begin{equation}
\nonumber
C = \frac{{\rho \pi }}{2}R_p^2V_v^2{\cos ^2}\left( {\psi  - \alpha } \right)\begin{array}{*{20}{l}}
{}&{\begin{array}{*{20}{l}}
{}&{{c_{{d_i}}}\left( {{\lambda _i},{\beta _i}} \right) = {A_i}\left( {{\lambda _i}} \right) + {B_i}\left( {{\lambda _i}} \right){\beta _i}}
\end{array}}
\end{array}
\end{equation}

\begin{equation}
\nonumber
{A_i}\left( {{\lambda _i}} \right)=a_0+a_1 {\lambda _i}+a_2{\lambda^2_i}+a_3{\lambda^3_i}
\begin{array}{*{20}{l}}
{}&{B_i}\left( {{\lambda _i}} \right)=b_0+b_1 {\lambda _i}+b_2{\lambda^2_i}+b_3{\lambda^3_i}
\end{array}
\end{equation}
with
$a_0=0.25382$, $a_1=-0.1369$, $a_2=0.04345$, $a3=-0.00263$, 
$b_0=-0.008608$, $b_1=0.0063$, $b_2=-0.0015$ and $b_3=0.000118$.\\
With repect to the fault, the vector field $\Lambda (x,t)$ and the decoupling matrix $\Theta (x,t)$ can be written as:
\begin{align}
\Lambda (x,t) = \left[ {\begin{array}{*{20}{l}}
{ - \frac{{{d_r}}}{{{f_r}}}\ddot \psi  + \frac{{B\left( \lambda  \right)C}}{{{f_r}{T_\beta }}}l\left( {{\beta _1} - {\beta _2}} \right) + \frac{{B\left( \lambda  \right)\dot C}}{{{f_r}}}l\left( {{\beta _1} - {\beta _2}} \right) + \frac{{\dot B\left( \lambda  \right)C}}{{{f_r}}}l\left( {{\beta _1} - {\beta _2}} \right)}\\
{{\Lambda _{11}}}\\
{\frac{1}{J}\left( {{\Gamma _{{a_1}}} - {\sigma _{11}}{\Lambda _{11}} - {\sigma _{12}}{\Lambda _{12}}} \right)}\\
{\sqrt {\frac{1}{3}} \left( {{f_{\bar\mu_5}} + {f_{\bar\mu_6}} + {f_{\bar\mu_7}}} \right)}\\
{{\Lambda _{21}}}\\
{\frac{1}{J}\left( {{\Gamma _{{a_2}}} - {\sigma _{21}}{\Lambda _{21}} - {\sigma _{22}}{\Lambda _{22}}} \right)}\\
{\sqrt {\frac{1}{3}} \left( {{f_{\bar\mu_9}} + {f_{\bar\mu_{10}}} + {f_{\bar\mu_{11}}}} \right)}
\end{array}} \right]
\end{align}
\begin{align}
&\nonumber{\Lambda _{11}} = \sqrt {\frac{2}{3}} \left( {\left[ {\begin{array}{*{20}{l}}
{\cos \left( {{\theta _{e1}}} \right)}~{\cos \left( {{\theta _{e1}} - \frac{{2\pi }}{3}} \right)}~{\cos \left( {{\theta _{e1}} + \frac{{2\pi }}{3}} \right)}
\end{array}} \right]{{\left[ {\begin{array}{*{20}{l}}
{{f_{\bar\mu_5}}}~{{f_{\bar\mu_6}}}~{{f_{\bar\mu_7}}}
\end{array}} \right]}^T} - p{\Omega _1}\left[ {\begin{array}{*{20}{l}}
{\sin \left( {{\theta _{e1}}} \right)}~{\sin \left( {{\theta _{e1}} - \frac{{2\pi }}{3}} \right)}~{\sin \left( {{\theta _{e1}} + \frac{{2\pi }}{3}} \right)}
\end{array}} \right]{I_1}} \right)\\\nonumber
&{\Lambda _{12}} =  - \sqrt {\frac{2}{3}} \left( {\left[ {\begin{array}{*{20}{l}}
{\sin \left( {{\theta _{e1}}} \right)}~{\sin \left( {{\theta _{e1}} - \frac{{2\pi }}{3}} \right)}~{\sin \left( {{\theta _{e1}} + \frac{{2\pi }}{3}} \right)}
\end{array}} \right]{{\left[ {\begin{array}{*{20}{l}}
{{f_{\bar\mu_5}}}~{{f_{\bar\mu_6}}}~{{f_{\bar\mu_7}}}
\end{array}} \right]}^T} - p{\Omega _1}\left[ {\begin{array}{*{20}{l}}
{\cos \left( {{\theta _{e1}}} \right)}~{\cos \left( {{\theta _{e1}} - \frac{{2\pi }}{3}} \right)}~{\cos \left( {{\theta _{e1}} + \frac{{2\pi }}{3}} \right)}
\end{array}} \right]{I_1}} \right)
\end{align}\\
\begin{align}
\nonumber
&{\Lambda _{21}} = \sqrt {\frac{2}{3}} \left( {\left[ {\begin{array}{*{20}{c}}
{\cos \left( {{\theta _{e2}}} \right)}~{\cos \left( {{\theta _{e2}} - \frac{{2\pi }}{3}} \right)}~{\cos \left( {{\theta _{e2}} + \frac{{2\pi }}{3}} \right)}
\end{array}} \right]{{\left[ {\begin{array}{*{20}{l}}
{{f_{\bar\mu_9}}}~{{f_{\bar\mu_{10}}}}~{{f_{\bar\mu_{11}}}}
\end{array}} \right]}^T} - p{\Omega _2}\left[ {\begin{array}{*{20}{l}}
{\sin \left( {{\theta _{e2}}} \right)}~{\sin \left( {{\theta _{e2}} - \frac{{2\pi }}{3}} \right)}~{\sin \left( {{\theta _{e2}} + \frac{{2\pi }}{3}} \right)}
\end{array}} \right]{I_2}} \right)\\
\nonumber
&{\Lambda _{22}} =  - \sqrt {\frac{2}{3}} \left( {\left[ {\begin{array}{*{20}{c}}
{\sin \left( {{\theta _{e2}}} \right)}~{\sin \left( {{\theta _{e2}} - \frac{{2\pi }}{3}} \right)}~{\sin \left( {{\theta _{e2}} + \frac{{2\pi }}{3}} \right)}
\end{array}} \right]{{\left[ {\begin{array}{*{20}{c}}
{{f_{\bar\mu_9}}}~{{f_{\bar\mu_{10}}}}~{{f_{\bar\mu_{11}}}}
\end{array}} \right]}^T} - p{\Omega _2}\left[ {\begin{array}{*{20}{c}}
{\cos \left( {{\theta _{e2}}} \right)}~{\cos \left( {{\theta _{e2}} - \frac{{2\pi }}{3}} \right)}~{\cos \left( {{\theta _{e2}} + \frac{{2\pi }}{3}} \right)}
\end{array}} \right]{I_2}} \right)
\end{align}
\begin{align}
&\nonumber{\sigma _{11}} = p{\phi _f} + \sqrt {\frac{2}{3}} p\left( {{L_d} - {L_q}} \right)\left[ {\begin{array}{*{20}{c}}
{\cos \left( {{\theta _{e1}}} \right)}~{\cos \left( {{\theta _{e1}} - \frac{{2\pi }}{3}} \right)}~{\cos \left( {{\theta _{e1}} + \frac{{2\pi }}{3}} \right)}
\end{array}} \right]{I_1}\\
\nonumber
&{\sigma _{12}} =  - \sqrt {\frac{2}{3}} p\left( {{L_d} - {L_q}} \right)\left[ {\begin{array}{*{20}{c}}
{\sin \left( {{\theta _{e1}}} \right)}~{\sin \left( {{\theta _{e1}} - \frac{{2\pi }}{3}} \right)}~{\sin \left( {{\theta _{e1}} + \frac{{2\pi }}{3}} \right)}
\end{array}} \right]{I_1}\\
\nonumber
&{\sigma _{21}} = p{\phi _f} + \sqrt {\frac{2}{3}} p\left( {{L_d} - {L_q}} \right)\left[ {\begin{array}{*{20}{c}}
{\cos \left( {{\theta _{e2}}} \right)}~{\cos \left( {{\theta _{e2}} - \frac{{2\pi }}{3}} \right)}~{\cos \left( {{\theta _{e2}} + \frac{{2\pi }}{3}} \right)}
\end{array}} \right]{I_2}\\
\nonumber
&{\sigma _{22}} =  - \sqrt {\frac{2}{3}} p\left( {{L_d} - {L_q}} \right)\left[ {\begin{array}{*{20}{c}}
{\sin \left( {{\theta _{e2}}} \right)}~{\sin \left( {{\theta _{e2}} - \frac{{2\pi }}{3}} \right)}~{\sin \left( {{\theta _{e2}} + \frac{{2\pi }}{3}} \right)}
\end{array}} \right]{I_2}
\end{align}\\
\begin{equation}
\Theta (x,t) = \left[ {\begin{array}{*{20}{c}}
{ - \frac{{2l}}{{{d_r}{T_\beta }}}BC}&0&0&0&0&0&0\\
0&{}&{}&{}&0&0&0\\
0&{}&{{P_1}{L^f}^{-1}}&{}&0&0&0\\
0&{}&{}&{}&0&0&0\\
0&0&0&0&{}&{}&{}\\
0&0&0&0&{}&{{P_2}{L^s}^{-1}}&{}\\
0&0&0&0&{}&{}&{}
\end{array}} \right]
\end{equation}\\
Noting that $P_i$ is the park transformation matrix:
\begin{equation}
P_i= \sqrt {\frac{2}{3}} \left[ {\begin{array}{*{20}{l}}
{\cos \left( {{\theta _{ei}}} \right)}&{\cos \left( {{\theta _{ei}} - \frac{{2\pi }}{3}} \right)}&{\cos \left( {{\theta _{ei}} + \frac{{2\pi }}{3}} \right)}\\
{ - \sin \left( {{\theta _{ei}}} \right)}&{ - \sin \left( {{\theta _{ei}} - \frac{{2\pi }}{3}} \right)}&{ - \sin \left( {{\theta _{ei}} + \frac{{2\pi }}{3}} \right)}\\
{\sqrt {\frac{1}{2}} }&{\sqrt {\frac{1}{2}} }&{\sqrt {\frac{1}{2}} }
\end{array}} \right]
\end{equation}
\end{strip}
\newpage
\quad
\end{document}